\documentclass[12pt]{article}
\usepackage[dvips]{graphics}
\usepackage{graphicx}
\begin{document}

\title{The Universal Arrow of Time}

\author{
Oleg Kupervasser$^1$\thanks{olegkup@yahoo.com},
Hrvoje Nikoli\'c$^2$\thanks{hrvoje@thphys.irb.hr},
Vinko Zlati\'c$^2$\thanks{Vinko.Zlatic@irb.hr}
\\
\\
\normalsize
$^1$ Scientific Research Computer Center,  Moscow State University,
Moscow, Russia \\
\normalsize
$^2$ Theoretical Physics Division, Rudjer Bo\v{s}kovi\'{c} Institute, \\
\normalsize
P.O.B. 180, HR-10002 Zagreb, Croatia
}

\maketitle

\begin{abstract}
Statistical physics cannot explain why a thermodynamic arrow of time exists, unless
one postulates very special and unnatural initial conditions. Yet, we argue that
statistical physics can explain
why the thermodynamic arrow of time is universal, i.e., why the arrow points
in the same direction everywhere.
Namely, if two subsystems have opposite arrow-directions 
at a particular time,
the interaction between them
makes the configuration statistically unstable and causes a decay towards a system
with a universal direction of the arrow of time.
We present general qualitative arguments for that claim
and support them by a detailed analysis of a toy model based
on the baker's map.
\end{abstract}

\vspace*{0.5cm}
PACS: 05.20.-y, 05.45.-a

Keywords: time arrow, entropy increase, baker's map

\section{Introduction}

The origin of the arrow of time is one of the greatest unsolved puzzles in physics
\cite{reichenbach,davies,penrose,price,zeh}.
It is well established that most arrows of time can be reduced to the
thermodynamic arrow, but the origin of the thermodynamic arrow
of time remains a mystery. Namely, the existence of the thermodynamic arrow
of time means that the system is not in the state with the highest possible entropy.
But this means that the system is not in the highest-probable state,
which lacks any statistical explanation. The fact that entropy increases with time
means that the system was in an even less probable state in the past, which makes the
problem even harder. Of course, the phenomenological fact
that entropy increases with time can be described by assuming that the universe
was in a state with a very low entropy at the beginning, but one cannot explain why
the universe started with such a very special and unnatural initial condition
in the first place.

Recently, Maccone \cite{mac} argued that the problem
of the origin of the arrow of time can be solved by quantum mechanics.
He has shown that in quantum mechanics all phenomena which leave a trail behind
(and hence can be studied by physics) are those the entropy of which increases.
(The observer's memory erasing  argument and the corresponding thought
experiments discussed in \cite{mac} has also been used previously for a resolution of entropy
increase and the quantum wave-packet reduction paradoxes \cite{vaidman,kuper04,kuper05}.)
From this he argued that the second law of thermodynamics is reduced to a mere
tautology, suggesting that it solves the problem of the arrow of time in physics.
However, several weaknesses on specific arguments used in \cite{mac}
have been reported \cite{com1,com2,com3}. As a response to one of these objections,
in a later publication  \cite{mac2} Maccone himself realized
that his approach does not completely solve the origin of the arrow of time
because the quantum mechanism he studied also requires highly improbable initial conditions
which cannot be explained.

Yet, as Maccone argued in \cite{mac2}, we believe that some ideas
presented in \cite{mac} and \cite{mac2} do help to better
understand the puzzle of the arrow of time. The purpose of this
paper is to further develop, refine, clarify, and extend some of
the ideas which were presented in \cite{mac,mac2,com2}, and also
in a somewhat different context in
\cite{kuper04,kuper05,zeh1,zeh2,kupervasser-big,thomson,Leb}. In
particular, unlike Maccone in \cite{mac,mac2}, we argue that
quantum mechanics is not essential at all. Indeed, in this paper
we consider only classical statistical mechanics.

The idea is the following. Even though statistical physics cannot explain why a thermodynamic arrow of time exists, we argue that at least it can explain
why the thermodynamic arrow of time is universal, i.e., why the arrow points
in the same direction everywhere.
Namely, if two subsystems have opposite arrow-directions initially,
we argue that the interaction between them
makes the configuration statistically unstable and causes a decay towards a system
with a universal direction of the arrow of time. This, of course, does not completely
resolve the problem of the origin of the arrow of time. Yet, at least,
we believe that this alleviates the problem.

The paper is organized as follows. In the next section we present
our main ideas in an intuitive non-technical form. After that, in
Sec.~\ref{SEC3} we study the statistical properties of the baker's
map (some basic properties of which are presented in the
Appendix), which serves as a toy model for studying generic
features of reversible chaotic Hamiltonian systems. As a
byproduct, in that section we also clarify the differences between
various notions of ``entropy''. Then, in Sec.~\ref{SEC4} we study
the effects of weak interactions between subsystems which, without
interactions, evolve according to the baker's map. In particular,
we explain how weak interactions destroy the opposite time arrows
of the subsystems, by making them much more improbable than
without interactions. Finally, in Sec.~\ref{SEC5} we present a
qualitative discussion of our results, including the consistency
with strongly-interacting systems in which the entropy of a
subsystem may decrease with time.

\section{Main ideas}
\label{SEC2}

To avoid ambiguity in further discussions, let us first explain our 
basic concepts and terminology used in the rest of the paper. 
For that purpose it is useful to visualize time as a continuous 1-dimensional
line. The line is coordinatized by a continuous parameter $t$. 
Such a coordinatization necessarily involves an orientation of the time-line
with a direction pointing from a smaller to a larger value of $t$.
We can use this orientation to define the notions such as 
``before'' and ``after'', ``past'' and ``future'', or ``initial'' and ``final''. 
In this paper, unless stated otherwise, by these notions we mean
the notions defined with respect to this coordinate time $t$.
However, we stress that such an orientation of coordinate time is merely
a matter of choice and does not have any physical content.
In particular, such an orientation by itself does {\em not}
involve any arrow of time. Instead, by an arrow of time we mean
a {\em physical} phenomenon, like a phenomenon that entropy
increases or decreases with time $t$. When the arrow of time points
in the same direction everywhere along the time-line, then the 
coordinate time can be defined such that the orientation of $t$
coincides with the arrow of time. 
That allows us to abuse the language at some places by arguing
that entropy ``increases'' rather than ``decreases'' with time, 
but one should have in mind that the difference between increase and decrease
with time is really a matter of definition. 
In general, 
the arrow of time and orientation of $t$ are logically independent concepts. 


Now let us discuss the thermodynamic arrow of time.
{\it A priori}, the probability of having a thermodynamic arrow of time is very low.
However, our idea is to think in terms of {\it conditional} probabilities.
Given that a thermodynamic arrow exists, what can we, by statistical arguments,
infer from that?

To answer that question, let us start from the laws of an underlying microscopic theory.
We assume that dynamics of microscopic degrees of freedom is described
by a set of second-order differential equations (with derivatives with respect to time)
which are invariant under the time inversion $t\rightarrow -t$. Thus, both directions
of time have an {\it a priori} equal roles. To specify a unique solution
of the dynamical equations of motion, one also needs to choose some ``initial'' time
$t_0$ on which initial conditions are to be specified. (The ``initial'' time does not necessarily need
to be the earliest time at which the universe came into the existence. For any $t_0$
at which the initial conditions are specified,
the dynamical equations of motion uniquely determine the state of the universe for both
$t>t_0$ and $t<t_0$.) It is a purely conventional particular instant on time,
which may be even in the ``future''.
Indeed, in this paper we adopt the ``block-universe'' picture of the world
(see, e.g., \cite{price,ellis,nik1,nik2} and references therein), according to which
time does not ``flow''. Instead, the universe is a ``static'' object extended in
4 spacetime dimensions.

Of course, the {\it a priori} probability of small entropy at $t_0$ is very low.
But {\em given that} entropy at $t_0$ is small, what is the conditional probability
that there is a thermodynamic arrow of time? It is, of course, very high.
However, given that entropy at $t_0$ is low, the most probable
option is that entropy increases in {\em both} directions with a minimum at $t_0$.
(We present an example in Fig.~\ref{figintsym} of Sec.~\ref{SECnumsim}.)
Clearly, in such a case the system is symmetric under the inversion 
$(t-t_0) \rightarrow - (t-t_0)$. The thermodynamic arrow of time for 
$t>t_0$ has the opposite direction than that for $t<t_0$.
Thus, even though neither direction of time has a preferred physical role
{\em globally}, a thermodynamic arrow of time can still be assigned {\em locally} for 
various times $t \neq t_0$.

On the other hand, at times at which we make measurements in practice,
the entropy is indeed low, but entropy does not increase in both directions.
Instead, it increases in only one direction. 
(In other words, a typical time $t_0$ at which we make a measurement
is not a time at which entropy attains a minimum, which is why nature
does not appear time-inversion symmetric to us.)
This is because, on a typical
$t_0$, not only the ``initial'' entropy is specified, but a particular direction
of the entropy increase is specified as well. 
At the microscopic level, this is related to the fact
that on $t_0$ one does not only need to specify the initial particle positions, but
also their initial velocities.


And now comes the central question of this section. Given that at $t_0$
the entropy is low, why entropy at $t_0$ increases in the same (say, positive)
direction everywhere? Isn't it more probable that the direction of entropy-increase
varies from point to point at $t_0$? If so, then why don't we observe it?
In other words, why the arrow of time is {\em universal}, having the same direction
everywhere for a given $t_0$? We refer to this problem as the problem
of {\em universality of the arrow of time}.

In this paper we argue that {\em this} problem can be solved by statistical physics.
In short, our solution is as follows. If we ignore the interactions between different degrees of
freedom, then, given that at $t_0$ the entropy is low, the most probable
option is, indeed, that the direction of the arrow of time varies from point to point.
On the other hand, if different degrees of freedom interact with each other,
then it is no longer the most probable option. Instead, even if the direction
of the arrow of time varies from point to point at $t_0$, the interaction provides
a natural mechanism that aligns all time arrows to the same direction.

To illustrate the arrow-of-time dilemma,
the thought experiments of Loschmidt (time reversal paradox) and Poincare (recurrence
theorem) are also often used. The corresponding paradoxes
in classical mechanics are resolved as follows. Classical mechanics allows, at
least in principle, to exclude any effect of the observer on the observed system.
However, most
realistic systems are {\em chaotic}, so –a weak perturbation may lead to an exponential divergence of trajectories.
In addition, there is also a non-negligible interaction.
As a simple example, consider a gas expanding from a small region of space into a large volume. In
this entropy-increasing process, the time evolution of macroscopic parameters is stable against
small external perturbations. On the other hand, if all the velocities are reversed, then the gas will
end up in the initial small volume, but only in the absence of any perturbations. The latter
entropy-decreasing process is clearly unstable and a small external perturbation would trigger
a continuous entropy growth. Thus the entropy increasing processes are stable, but the
decreasing ones are not. A natural consequence is that the time arrows (the directions of which are
defined by the entropy growth) of both the observer and the observed system are aligned
to the same direction,
because of the inevitable non-negligible interaction between them. They can return back to the
initial state only together (as a whole system) in both Loschmidt and Poincare paradoxes.
so the observer’'s memory gets erased in the end. 
In particular, the observer’'s memory gets erased in the end,
because we assume that everything, including the observer's brain, 
is reversed to the state identical to an old state describing the system
before the memory has been created.
During this process
the time arrow of the observer points in the backward direction, which has two consequences.
First, an entropy growth is observed in the whole system as well as in its two parts,
despite the fact that entropy decreases with coordinate time.
Second, the memory of the observer is erased not only at the end but
also close to that point, because the observer does not remember his ``“past''”
(defined with respect to the coordinate time), but remembers his ``“future''”.
Of course, the observer himself cannot know that the arrow of time has reversed its
direction, because he can only observe the {\em physical}  
``“past''” and ``“future'' defined not with respect to the coordinate time, but
with respect to the direction in which entropy increases.

Indeed, it may seem quite plausible that interaction will align all time arrows to the same
direction. But the problem is - which direction? The forward direction,
or the backward one? How can any particular direction be preferred,
when both directions are {\it a priori} equally probable?
Is the common direction chosen in an effectively random
manner, such that it cannot be efficiently predicted? Or if there are two subsystems
with opposite directions of time at $t_0$,
will the ``stronger'' subsystem (i.e., the one with a larger number of degrees of freedom)
win, such that the joint system will take the direction of the
``stronger'' subsystem as their common direction?

The answer is as follows:
It is all about conditional probabilities. One cannot question the facts which are
already known, irrespective of whether these facts are in ``future'' or ``past''.
The probabilistic reasoning is to be applied to only those facts which are not known yet.
So, let us assume that the entropy is low at $t_0$ and that we have two subsystems
with opposite time directions at $t_0$. Let us also assume that
the subsystems do not come into a mutual interaction before $t_1$ (where $t_1>t_0$),
after which they interact with each other.
Given all that, we {\em know} that, for $t_0<t\leq t_1$, entropy increases with time
for one subsystem and decreases with time for another subsystem.
But what happens for $t>t_1$? Due to the interaction, the two subsystems will have
the same direction of the arrow of time for $t>t_1$. But which direction? The probabilistic answer is: The direction which is more probable, {\em given} that we know what we already know. But we already know
the situation for $t<t_1$ (or more precisely, for $t_0<t\leq t_1$),
so our probabilistic reasoning can only be applied to $t>t_1$.
It is this {\em asymmetry in knowledge} that makes two directions of time
different.
(Of course, the interaction is also asymmetric, in the sense that
interaction exists for $t>t_1$, but not for $t_0<t\leq t_1$.)
Thus, the probabilistic reasoning implies that entropy will increase
in the {\em positive} time direction for $t>t_1$.
Alternatively, if there was no such asymmetry in knowledge, we could not
efficiently predict the direction of the arrow of time, so the joint direction
would be chosen in an effectively random manner.

Note also that the mechanism above does not depend much on relative sizes 
of the two subsystems. In particular, if they are of an equal size,
i.e., half of the system has the time arrow oriented opposite
to the other half for $t_0<t\leq t_1$, the common time arrow 
for $t>t_1$ will still be determined by the aforementioned asymmetry
in knowledge.

Further note that the qualitative probabilistic arguments above refer to 
expectations on typical systems, not necessarily on all possible systems.
Indeed, there are interesting physical systems, such as those involving the spin-echo effect,
in which a subsystem may have the arrow of time opposite to that of its
environment. The point is that such systems are exceptions, rather than a rule.
Thus, our qualitative probabilistic arguments against such systems
are still justified, provided that one does not misinterpret them as strict laws
without exceptions.

In fact, it is not difficult to understand qualitatively why exceptions such 
as the spin-echo effect exist. First, it is a system with a relatively small number of 
degrees of freedom, which makes the statistical arguments less accurate
and fluctuations more likely. Second, the interaction of this system 
with the environment is so weak so that the mechanism of entropy alignment takes
more time than in most other systems. Indeed, even the spin-echo system
will eventually, after a sufficient time, align its time-direction with that of its
environment.

To emphasize that no direction of time is {\it a priori} preferred,
let us also briefly discuss a situation inverse to the above. For that
purpose, now let us assume that interaction exists only for $t<t_{-1}$,
where $t_{-1}<t_0$. By completely analogous reasoning, now we can conclude
that the entropy will increase in the {\em negative} time direction for
$t<t_{-1}$. 

A fully symmetric variant is also possible by assuming that the interaction exists
for both $t>t_1$ and $t<t_{-1}$ (but not for $t_{-1}\leq t \leq t_1$).
In this case, the entropy will increase
in the {\em positive} time direction for $t>t_1$ {\em and}
in the {\em negative} time direction for $t<t_{-1}$. 
In other words, similarly to the case in Fig.~\ref{figintsym} of Sec.~\ref{SECnumsim},
the entropy will increase in {\em both} directions, but for different times.

One additional note on initial conditions is in order. Even when two subsystems
have the opposite directions of the thermodynamic arrow of time, 
we choose the initial conditions for both of them to be at the {\em same} time $t_0$,
say in the past. Indeed, the choice that $t_0$ is in the past is a natural
choice for the subsystem in which entropy increases with time. However, this choice
is not so natural for the other subsystem in which  entropy decreases with time;
for that subsystem it would be more natural to choose the ``initial'' condition
in the future. Or more generally, one might like to study many subsystems,
each with initial conditions at another time. We stress that in this paper
we do not study such more general initial conditions because, when 
interactions between the subsystems are present, initial conditions at different times
for different subsystems cannot be chosen arbitrarily. Namely, for some choices
of such initial conditions at different times, a self-consistent solution of the 
dynamical equations of motion may not even exist. And even when a solution exists,
in general it is not known how to prove the existence or how to
find a solution numerically.

Now we can understand why the arrow of time appears to be universal. If there is
a subsystem which has an arrow of time opposite to the time-arrow that we are used to,
then this subsystem is either observed or not observed by us.
If it is not observed, then it does not violate the fact that the arrow of time appears
universal to us. If it is observed then it interacts with us, which implies that it cannot
have the opposite arrow of time for a long time. In each case,
the effect is that {\em all what we observe must have the same direction of the
arrow of time}
(except, perhaps, during a very short time interval).
This is similar to the reasoning in \cite{mac}, with an important difference that our
reasoning does not rest on quantum mechanics.

In the remaining sections we support these intuitive ideas by a more quantitative
analysis.

\section{Statistical mechanics of the baker's map}
\label{SEC3}

The baker's map (for more details see Appendix \ref{app1})
maps any point of the unit square to another point of the same square.
We study a collection of $N \gg 1$ such points (called ``particles'')
that move under the baker's map. This serves as a toy model for a ``gas'' that shares
all typical properties of classical Hamiltonian reversible deterministic chaotic systems.
Indeed, due to its simplicity, the baker's map is widely used for such purposes
\cite{prigogine1,elskens,gaspard,hartmann}.

\subsection{Macroscopic entropy and ensemble entropy}

To define a convenient set of macroscopic variables, we divide the unit square into 4 equal subsquares. Then
the 4 variables $N_1$, $N_2$, $N_3$, $N_4$, denoting the number of
``particles'' in the corresponding subsquares, are defined to be the macroscopic variables
for our system. (There are, of course, many other convenient ways to define macroscopic
variables, but general statistical conclusions are not expected to depend on this choice.)
The {\em macroscopic entropy} $S_{\rm m}$ of a given macrostate
is defined by the number of different microstates corresponding to that macrostate, as
\begin{equation}\label{macrent}
 S_{\rm m}=-N\sum_{k=1}^{4} \frac{N_k}{N} \, {\rm log} \left( \frac{N_k}{N} \right)
=- \sum_{k=1}^{4} N_k \, {\rm log} \left( \frac{N_k}{N} \right) .
\end{equation}
This entropy is maximal when the distribution of particles is uniform, in which
case $S_{\rm m}$ is $S_{\rm m}^{\rm max}=N \, {\rm log} 4$. Similarly, the entropy is minimal
when all particles are in the same subsquare, in which case $S_{\rm m}=0$.

Let $(x,y)$ denote the coordinates of a point on the unit square.
In physical language, it corresponds to the particle position
in the 2-dimensional phase space. For $N$ particles, we consider a statistical ensemble
with a probability density $\rho(x_1,y_1; \ldots ;x_N,y_N;t)$
on the $2N$ dimensional phase space. Here $t$ is the evolution
parameter, which takes discrete values $t=0,1,2,\ldots$ for the baker's map.
Then the {\em ensemble entropy} is defined as
\begin{equation}
 S_{\rm e}=-\int \rho(x_1,y_1; \ldots ;x_N,y_N;t) \; \log \rho(x_1,y_1; \ldots ;x_N,y_N;t) \, dX ,
\end{equation}
where
\begin{equation}
 dX \equiv dx_1 \, dy_1 \, \cdots dx_N \, dy_N.
\end{equation}
In general, $\rho$ and $S_{\rm e}$ change during the evolution generated by the
baker's map and depend on the initial $\rho$.
However, if the initial probability-density function has a form
\begin{equation}
 \rho(x_1,y_1; \ldots ;x_N,y_N) = \rho(x_1,y_1) \cdots \rho(x_N,y_N) ,
\end{equation}
which corresponds to an uncorrelated density function, then
the probability-density function remains uncorrelated during the evolution.

As an example, consider $\rho(x_l,y_l)$ which is uniform within some
subregion $\Sigma$ (with area $A<1$) of the unit square,
and vanishes outside of $\Sigma$. In other words, let
\begin{equation}
 \rho(x_l,y_l,t)=
\left\{
\begin{array}{l}
\displaystyle
1/A \;\;\;\; {\rm for} \; (x_l,y_l) \;  {\rm inside}\; \Sigma, \\
\displaystyle
0  \;\;\;\; {\rm for} \; (x_l,y_l) \;  {\rm outside}\; \Sigma .
\end{array}
\right.
\end{equation}
In this case
\begin{equation}
 S_{\rm e}=-\left(\frac{1}{A}\right)^N  {\rm log} \left(\frac{1}{A}\right)^N  A^N
= N \, {\rm log} A.
\end{equation}
Since $A$ does not change during the baker's map evolution, we see that
$S_{\rm e}$ is constant during the baker's map evolution. This example
can be used to show that $S_{\rm e}$ is, in fact,
constant for {\em arbitrary} initial probability function. To briefly sketch the proof,
let us divide the unit $2N$-dimensional box into a large number of small regions $\Sigma_a$,
on each of which the probability is equal to $\rho_a$.
During the evolution, each region $\Sigma_a$ changes the shape, but its
$2N$-dimensional ``area'' $A_a$ remains the same. Moreover, the probability $\rho_a$ on the
new $\Sigma_a$ also remains the same. Consequently,
the ensemble entropy $S_{\rm e}=-\sum_a  A_a^N \rho_a \, {\rm log}\rho_a$
remains the same as well. This is the basic idea
of a discrete version of the proof, but a continuous version can be done in a similar way.

\subsection{Appropriate and inappropriate macroscopic variables}

The macroscopic variables defined in the preceding subsection have the following
properties:
\begin{enumerate}
 \item For most initial microstates having the property $S_{\rm m}<S_{\rm m}^{\rm max}$,
$S_{\rm m}$ increases during the baker's map.
 \item For most initial microstates having the property $S_{\rm m}=S_{\rm m}^{\rm max}$,
$S_{\rm m}$ remains constant during the baker's map.
 \item The two properties above do not change when the baker's map
is perturbed by a small noise.
\end{enumerate}
We refer to macrovariables having these properties as {\em appropriate} macrovariables.
(They are ``appropriate'' in the sense that an appropriate macroscopic law of entropy-increase
can be obtained only when macrovariables obey these properties.)

Naively, one might think that any seemingly reasonable choice of macrovariables
is appropriate. Yet, this is not really the case. Let us demonstrate this
by an example. Let us divide the unit square into $2^M$ equal vertical strips ($M \gg 1$).
We define a new set of macrovariables as the numbers of
particles inside each of these strips.
Similarly to (\ref{macrent}), the corresponding macroscopic entropy is
\begin{equation}\label{macrent2}
 S_{\rm m}=- \sum_{k=1}^{2^M} N_k \, {\rm log} \left( \frac{N_k}{N} \right) ,
\end{equation}
where $N_k$ is the number of particles in strip $k$.
For the initial condition, assume that
the gas is uniformly distributed inside odd vertical strips, while even strips are empty. Then
$S_{\rm m}<S_{\rm m}^{\rm max}$ initially. Yet, for a long time during the baker's evolution,
$S_{\rm m}$ does not increase for any initial microstate corresponding to this macrostate.
However, during this evolution the number of
filled strips decreases and their thickness increases, until only one thick filled vertical strip
remains. After that, $S_{\rm m}$ starts to increase.
We also note that the evolution towards the single strip can be easily destroyed
by a small perturbation.

Thus we see that vertical strips lead to inappropriate macrovariables.
By contrast, horizontal strips lead to appropriate macrovariables.
(Yet, the macrovariables in (\ref{macrent}) are even more appropriate,
because they lead to much faster growth of  $S_{\rm m}$.)
This asymmetry between vertical and horizontal strips is a consequence
of the intrinsic asymmetry of the baker's map with respect to vertical
and horizontal coordinates. This asymmetry is analogous to the asymmetry
between canonical coordinates and momenta in many realistic Hamiltonian systems
of classical mechanics. Namely, most realistic Hamiltonian systems contain
only local interaction between particles, where locality refers to a separation
in the coordinate (not momentum!) space.

Finally, we note that evolution of the macroscopic variables $N_k(t)$, $k=1,2,3,4$,
is found by averaging over ensemble in the following way
\begin{equation}
 N_k(t)=\int N_k(x_1,y_1; \ldots ;x_N,y_N;t)    \rho(x_1,y_1; \ldots ;x_N,y_N;t)  \, dX .
\end{equation}

\subsection{Coarsening}
\label{SECcoars}

As we have already said, the ensemble entropy (unlike macroscopic entropy)
is always constant during the baker's map evolution.
One would like to have a modified definition of the ensemble entropy that
increases similarly to the macroscopic entropy. Such a modification is
provided by {\em coarsening}, which can be defined by introducing
a coarsened probability-density function
\begin{eqnarray}
\rho^{\rm coar}(x_1,y_1; \ldots ;x_N,y_N) & = &
\int \Delta(x_1-x'_1,y_1-y'_1; \ldots ;x_N-x'_N,y_N-y'_N)
\nonumber \\
& & \times \, \rho(x'_1,y'_1; \ldots ;x'_N,y'_N) \, dX' ,
\end{eqnarray}
 where $\Delta$ is nonvanishing in some neighborhood of
$X'=0,0;\ldots ;0,0$.
In this way, the coarsened ensemble entropy is
\begin{equation}
 S_{\rm e}^{\rm coar}=-\int \rho^{\rm coar}(x_1,y_1; \ldots ;x_N,y_N) \;
\log \rho^{\rm coar}(x_1,y_1; \ldots ;x_N,y_N) \, dX .
\end{equation}
Of course, the function $\Delta$ can be chosen in many ways. In the following we discuss
a few examples.

One example is the Boltzmann coarsening, defined by
\begin{equation}
\rho^{\rm coar}(x_1,y_1; \ldots ;x_N,y_N) = \rho(x_1,y_1) \cdots \rho(x_N,y_N) ,
\end{equation}
where
\begin{equation}
 \rho(x_1,y_1) = \int \rho(x_1,y_1; \ldots ;x_N,y_N) \,
dx_2 \, dy_2 \, \cdots dx_N \, dy_N ,
\end{equation}
and similarly for other $\rho(x_l,y_l)$.

Another example is isotropic coarsening, having a form
\begin{eqnarray}
& \Delta(x_1-x'_1,y_1-y'_1; \ldots ;x_N-x'_N,y_N-y'_N) = &
\nonumber \\
& \Delta(x_1-x'_1) \Delta(y_1-y'_1) \cdots \Delta(x_N-x'_N) \Delta(y_N-y'_N) . &
\end{eqnarray}

Yet another example is the Prigogine coarsening \cite{prigogine1}
\begin{equation}
\Delta(x_1-x'_1,y_1-y'_1; \ldots ;x_N-x'_N,y_N-y'_N) =
\Delta(y_1-y'_1)  \cdots  \Delta(y_N-y'_N) ,
\end{equation}
which is an anisotropic coarsening over the shrinking direction $y$.

Finally, let us mention the coarsening based on dividing the system into
two smaller interacting subsystems. The coarsened ensemble entropy for the full system
is defined as the sum of uncoarsened ensemble entropies of its subsystems.
Such a coarsened entropy ignores the correlations between the subsystems.

All these types of coarsening have the following property:
If the initial microstate is such that macroscopic entropy increases, then
the coarsened ensemble entropy also increases for that initial microstate.
Yet, the Prigogine coarsening has the following advantages over
Boltzmann and isotropic coarsenings:

First, if the initial microstate is such that the macroscopic entropy {\em decreases},
then the Prigogine coarsened ensemble entropy does {\em not} decrease,
while the Boltzmann and isotropic coarsened ensemble entropies decrease.

Second, assume that the initial microstate is such that the
macroscopic entropy increases, and consider some ``final'' state
with a large macroscopic entropy close to the maximal one. After
this final state, consider the corresponding ``inverted'' state, (i.e., the
state with exchanged $x$ and $y$). In the transition to this
``inverted'' state, the Prigogine coarsened ensemble entropy decreases, while the
Boltzmann and isotropic coarsened ensemble entropies remain
unchanged.

Thus, the Prigogine coarsening provides the most correct description of the
ensemble-entropy increase law without any additional assumptions.
For example, to get the same result with Boltzmann coarsening, one would need
to introduce the additional ``molecular chaos hypothesis'' to replace
$\rho(x_1,y_1;x_2,y_2)$ with $\rho(x_1,y_1)\rho(x_2,y_2)$
in the equation of motion for $\rho(x,y,t)$.

\section{The effects of weak interactions}
\label{SEC4}

\subsection{Small external perturbations}

The growth of the ensemble entropy can be achieved even without
coarsening, by introducing a small external perturbation of the baker's map.
The perturbation must be small enough to avoid destruction of the
growth of macroscopic entropy, but at the same time, it must be
strong enough to destroy the reverse processes and Poincare returns.
For most such perturbations, the qualitative features of the evolution
do not depend much on details of the perturbation.

There are two ways how the external perturbation can be
introduced. One way is to introduce a small external random noise.
The macroscopic processes with the increase of macroscopic entropy
are stable under such a noise. However, the area of a region is no
longer invariant under the perturbed baker's map. In this way the
ensemble entropy can increase.

The other way is to introduce a weak interaction with the
environment (which can be thought of as an ``observer''). Again,
the macroscopic processes with the increase of macroscopic entropy
are stable, but the area of a region is no longer invariant under
the perturbed baker's map. Consequently, the ensemble entropy can
increase. However, such a system is no longer isolated. Instead,
it is a part of a larger system divided into two subsystems.
Hence, as we have already explained in Sec.~\ref{SECcoars}, the
coarsened ensemble entropy for the full system can be defined as
the sum of uncoarsened ensemble entropies of its subsystems. In
the next subsection we study the weak interactions with the
environment in more detail.

\subsection{Weak interaction and the destruction of opposite time arrows}
\label{SECweakint}

To proceed, one needs to choose some specific interaction between
two gases. In the absence of interaction, each of them evolves
according to the baker's map. We put the two unit squares one
above another and specify the interaction with the range of interaction $\sigma$,
such that, between two steps of the baker'’s map, all closest
pairs of particles (with distance smaller than $\sigma$ between
them) exchange their positions. (More precisely, we first find the
pair of closest particles (with distance smaller than $\sigma$
between them) and exchange their positions. After that, we find
the second pair of closest particles (with distance smaller than
$\sigma$ between them and different from previously chosen
particles) and exchange their positions too. We repeat this
procedure until we exhaust all particles.) The interaction happens
{\em only} between the particles in different subsystems because 
it makes no sense to introduce such interaction within a single
subsystem. Namely, 
such an interaction does not affect the motion of
the particles, but only gives rise to the mixing between the two
subsystems when two particles of the pair belong to different
subsystems. 
When they belong to the same system, we interpret them
as trivial irrelevant exchanges, and consequently think of them as
exchanges that have not happened at all.

Note also that such mixing by itself does not lead to the Gibbs
paradox, as long as we consider the two unit squares as separate
objects. The macroscopic entropy is defined as the sum of
macroscopic entropies of the two subsystems.

Now let us consider the case in which the time arrows of the two subsystems
have the same direction.
The processes in which the macroscopic entropies of the two subsystems increase
are stable under the interaction. Thus, most low-entropy initial conditions
lead to a growth of macroscopic entropy of both subsystems, as well as of
the full system.

Similarly, if we inverse a process above with increasing macroscopic entropy,
we obtain a system in which
macroscopic entropy of both subsystems, as well as of
the full system - {\em decreases}. In this sense, the interaction does not ruin the symmetry between
the two directions of time.

Now let us consider the most interesting case, in which entropy increases
in the first subsystem and decreases in the second. The initial state
of the first subsystem has a low entropy (for example, all particles
are in some small square near the point $(0,0)$ of the unit square).
Likewise, the second system has a low entropy
(for example, all particles
are in some small square near the point $(1,1)$ of the unit square)
in the {\em final} state.

If there was no interaction, the final state of the first subsystem
would be a high-entropy state corresponding to a nearly uniform
distribution of particles. Likewise, the initial state of the second system
would be a high-entropy state of the same form.

However, the solutions above with two opposite arrows of time
are no longer solutions when the interaction is present.
In most cases, the interaction mixes the particles between the subsystems.
The number of solutions with interaction which have the initial-final
conditions prescribed above is very small, in fact {\em much smaller than
the number of such solutions in the absence of interaction.}

Let us make the last assertion more quantitative.
After an odd number of (non-trivial) exchanges, the particle
transits to the other subsystem. Likewise, after
an even number of such exchanges, it remains in the same subsystem.
The probabilities for these two events are equal
to $p=1/2$ and do not depend on other particles, at least
approximately.
Further, we can argue that the mixing between the two
subsystems is negligible in the initial and final states,
as the entropies of the two subsystems are very different.
We want to calculate the probability of a small mixing
in the final state, given that the mixing is small in the
initial state. For definiteness, we shall say that
the mixing is small if the number $N_t$ of transited particles
is either $N_t<N/4$ or $N_t>3N/4$.
(Recall that the particle exchange is the {\em only} effect of our perturbation,
so mixing cannot depend on any other variable except $N_t$.
The factors $1/4$ and $3/4$ are, of course, chosen arbitrarily, but we fix
them in order to get some concrete results in a simple final form.
The qualitative results that follow do not depend on this choice.)
Thus, the probability is given by the cumulative
binomial distribution $F(N_t;N,1/2)$, given by
\begin{equation}
 F(k;n,p) = \sum_{i=0}^{\lfloor k\rfloor}
\left(
\begin{array}{l}
n \\ i
\end{array}
\right)
p^i (1-p)^{n-i}
\end{equation}
where $\lfloor k\rfloor$ is the greatest integer less than or equal to $k$.
The function $F(k;n,p)$ satisfies the bound
\begin{equation}
 F(k;n,p) \leq \exp \left( -2\frac{(np-k)^2}{n} \right) .
\end{equation}
Thus, since the opposite time arrows of subsystems are not destroyed
when $N_t<N/4$ or $N_t>3N/4$, we see that the probability of this
is equal to
\begin{equation}
 2F(N/4;N,1/2) \leq 2e^{-N/8} .
\end{equation}
Clearly, it decreases exponentially with $N$, which means that such a probability
is negligibly small for large $N$.
Hence, it is almost certain that processes with opposite time arrows will be destroyed.

In the model above, we need a nearly equal number of particles
in the two subsystems to destroy the opposite time arrows.
This is because one particle can influence the motion of only one
close particle. For more realistic interactions,
one particle can influence the motion of a large number of particles
in its neighborhood, which means that even a very small number of
particles in one system can destroy the entropy decreasing processes of the
other system.

\subsection{Decorrelation in the interacting system}

Hamiltonian systems are described not only by a macrostate, but also
by complex nonlinear correlations between microstates. These correlations are responsible for
reversibility. The interaction between two subsystems destroys these correlations inside the
subsystems, but the full system remains reversible, i.e.,
the correlations appear in the full system. Thus, the decorrelation in the subsystems
spreads the correlations over the full system. (This process is a classical
analogue of decoherence in quantum mechanics.)

Let us put these qualitative ideas into a more quantitative form.
Linear (Pearson) correlations
have a behavior very similar to the nonlinear correlations described above. The only
difference is that these linear correlations decrease with time.
The interaction we proposed can be approximated by a random noise with
amplitude corresponding to the range of interaction $\sigma$ between
the particles.
Therefore, we expect that
the interaction not only causes the alignment of the time arrows, but also
a decay of correlation which is even stronger than that without the interactions
(Sec.~\ref{SECdecor}). During this process the evolution of subsystems is irreversible,
but the full system remains reversible.

We can quantify this decay of correlations by calculating the Pearson correlation for our
subsystems, given by
\begin{equation}
 r(m)=\frac{C(m)}{\sqrt{C(0)\langle C^m(0)\rangle}},
\end{equation}
where $\langle C^m(0)\rangle$ is the expected variance of the random variable $x$ calculated
after $m$ iterations of the map. The variance $C^m(0)$ can be calculated as
\begin{equation}
 C^m(0)=\sum_{j=0}^{2^m-1}\int\limits_{j2^{-m}}^{(j+1)2^{-m}}\left(2^mx-j-\langle
x\rangle+S\right)^2dx,
\end{equation}
where $S$ is a random number defined as $S=\sum_{k=0}^{m-1}2^k\zeta_k$. Here $\zeta_k$
is an i.i.d. random number with zero mean
and variance $\sigma^2$, which models the influence of interactions on the evolution of
the system. After a short calculation we get
\begin{equation}
 \langle C^m(0)\rangle=C(0)+\langle S^2\rangle=C(0)+\sum_{k,k'=0}^{m-1}2^{k+k'}\langle
\zeta_k\zeta_{k'}\rangle .
\end{equation}
Using the properties of i.i.d. variables
$\langle\zeta_k\zeta_{k'}\rangle=\delta_{kk'}\sigma^2$,  it follows that
\begin{equation}
  \langle C^m(0)\rangle=C(0)+\frac{2^{2m}-1}{3}\sigma^2.
\end{equation}
It is clear that the interactions will enhance the decay of correlations of at least linear
dependencies, because
\begin{equation}\label{eq22}
 r(m)=\frac{2^{-m}}{\sqrt{1+4(2^{2m}-1)\sigma^2}}.
\end{equation}
Yet, for the full system the Pearson correlation $r(m)=2^{-m}$ remains the same.
Since $\langle S^2 \rangle^{1/2}$ must be much smaller than the system size
(unit square), we can conclude that
our assumptions resulting in (\ref{eq22}) are correct only for
$\langle S^2 \rangle = [(2^{2m}-1)/3]\sigma^2 \ll 1$ and
$ \sigma^2/2^{-2m}\ll 1$.


\subsection{Numerical simulation}
\label{SECnumsim}

So far, we have been using general abstract arguments. In this
subsection we support these arguments by a concrete numerical simulation.
We consider two subsystems (labeled as 1 and 2), each with
$N_1=N_2=300$ particles. The two subsystems occupy two unit squares.
To define the coarsened entropy,
each unit square is divided into $16\times 16 =256$ small squares.
Thus, the entropy in the two subsystems is given by
\begin{equation}
 S_{i}=-N_i \sum_{k=1}^{512} f_{k,i} \log f_{k,i},
\end{equation}
where $i=1,2$, $f_{k,i}=n_{k,i}/N_i$, and $n_{k,i}$ is the number of
particles in the corresponding small square.
Similarly, the total entropy is defined as
\begin{equation}
 S=-(N_1+N_2)\sum_{k=1}^{512} f_{k} \log f_{k},
\end{equation}
where $f_{k}=(n_{k,1}+n_{k,2})/(N_1+N_2)$

\begin{figure}[t]
\centerline{\includegraphics[width=8cm]{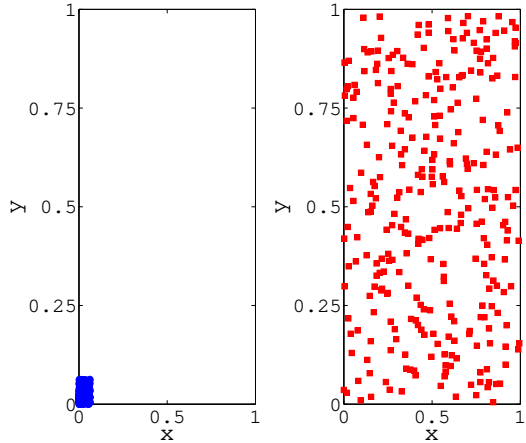}}
\caption{\label{t=1}The initial particle configuration at $t=1$.}
\end{figure}

\begin{figure}[t]
\centerline{\includegraphics[width=8cm]{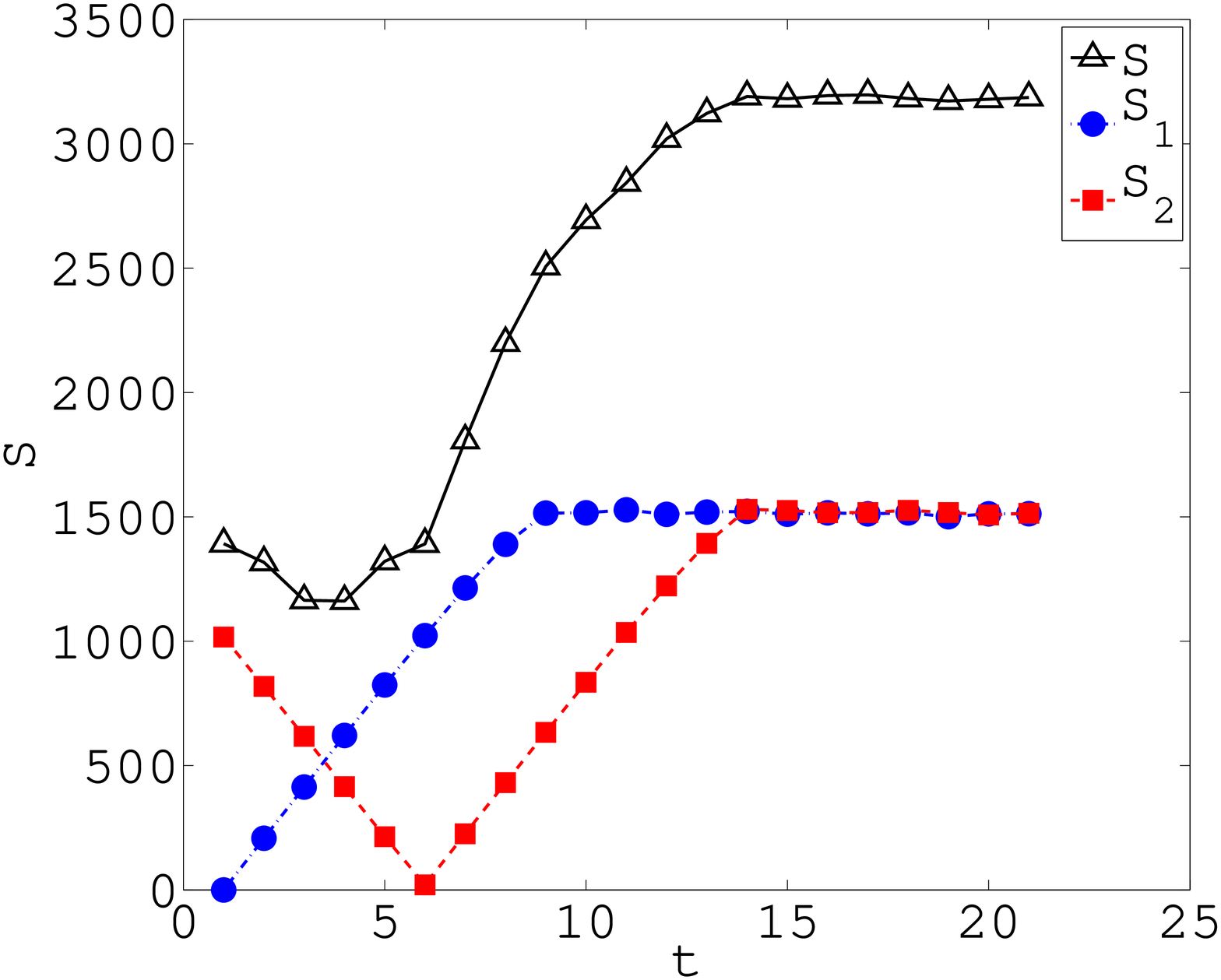}}
\caption{\label{fignoint}Evolution of entropy without interaction.}
\end{figure}

For the system 1 we choose a zero-entropy initial state at $t=1$
(see Fig.~\ref{t=1}).
Similarly, for the system 2 we choose a zero-entropy ``initial'' state at $t=6$.
Such initial conditions provide that, in the absence of interactions,
$S_1$ increases with time, while $S_2$ decreases with time for $t< 6$.
To avoid numerical problems arising from the finite precision
of computer representation of rational numbers,
(\ref{eq27}) is replaced by $x' = ax - \lfloor ax\rfloor$, $y'=(y+\lfloor ax\rfloor)/2$,
with $a=1.999999$.
The results of a numerical simulation are presented in
Fig.~\ref{t=1} and Fig.~\ref{fignoint}.

\begin{figure}[t]
\centerline{\includegraphics[width=8cm]{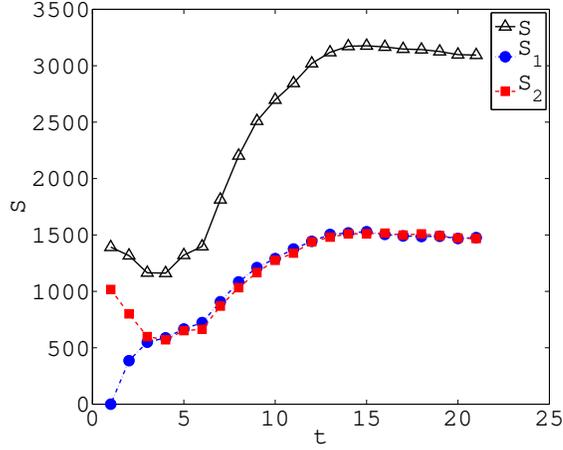}}
\caption{\label{figint} Evolution of entropy with interaction.}
\end{figure}

To include the effects of interaction, we define interaction in the following way.
(For the sake of computational convenience, it is defined slightly differently
than in Sec.~\ref{SECweakint}.
The interaction used in Sec.~\ref{SECweakint} is chosen to make the analytical
calculations simpler, while that in the present section is chosen to make the 
numerical computations simpler.)
We take a small range of interaction
$r_y=0.01$ in the $y$-direction, which can be thought of as a parameter that measures the weakness of interaction. (Recall that $y$ and $x$ are analogous to a canonical coordinate
and a canonical momentum, respectively, in a Hamiltonian phase space.)
The interaction exchanges the closest pairs similarly as in Sec.~\ref{SECweakint},
but now ``the closest'' refers to the distance in the $y$-direction, and there is no
exchange if the closest distance is larger than $r_y$.
In addition, now interaction is defined such that only the $x$-coordinates of the particles
are exchanged.
By choosing the same initial conditions at $t=1$ as in the non-interacting case
(Fig.~\ref{t=1}),
the results of a numerical simulation with the interaction are presented in Fig.~\ref{figint}.
We see that with interaction (Fig.~\ref{figint}) $S_2$ starts to increase at earlier times than without interaction (Fig.~\ref{fignoint}).

\begin{figure}[t]
\centerline{\includegraphics[width=8cm]{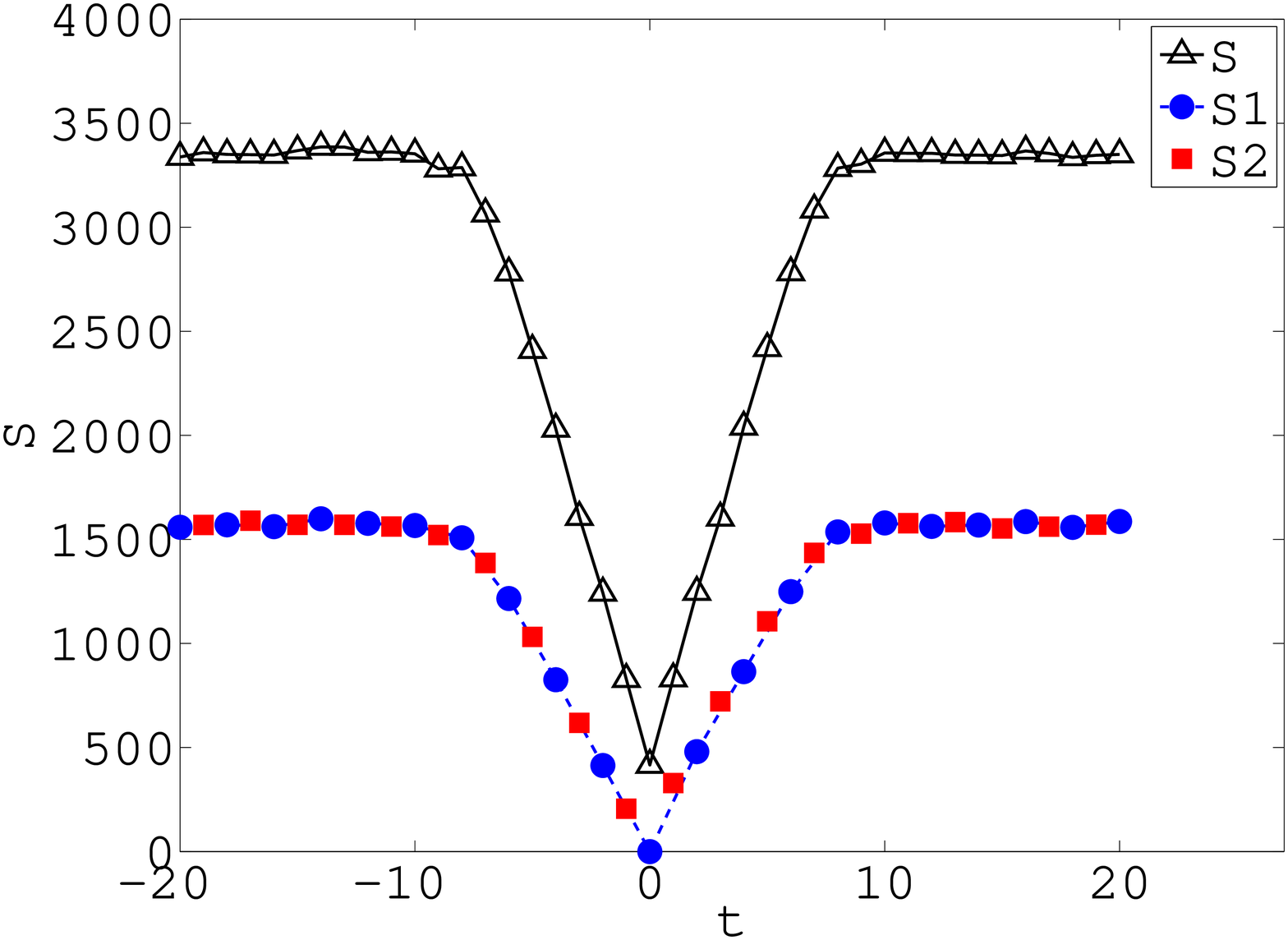}}
\caption{\label{figintsym} A symmetric evolution of entropy with interaction.}
\end{figure}

Finally, we present an example of a symmetric time evolution with interaction. For that purpose,
we choose a zero-entropy initial state at $t=0$ for both system 1 and system 2.
The two initial zero-entropy configurations are different. In particular, they
are located within two different small squares,
so that the total initial entropy is larger than zero.
The result of a numerical simulation is presented in Fig.~\ref{figintsym}.
We see that the solution is symmetric under the time inversion 
$t \rightarrow - t$.

\section{Discussion}
\label{SEC5}

In this paper, we have used the toy model based on the baker's map
to demonstrate features which seem to be valid for general systems
described by reversible Hamiltonian mechanics. Clearly, for such
systems one can freely choose either final or initial conditions,
but one cannot freely choose mixed initial-final conditions. (For mixed
initial-final conditions, the canonical variables
are fixed at an initial time for {\em some} of the particles,
but at an final time for the {\em other} particles.
For most such mixed initial-final conditions, an
appropriate solution (of the Hamiltonian equations of motion) does
not exist. Similarly, our toy model suggests that for most
Hamiltonians with weak interactions, the number of solutions with
given coarse-grained initial-final conditions is much smaller then
the number of solutions with only coarse-grained initial or only
coarse-grained final conditions. This explains why, in practice,
we almost never observe subsystems with opposite arrows of time, i.e.,
why the arrow of time is universal.

In a sense, this destruction of opposite arrows of time
is similar to ergodicity. Both properties are valid
for most practical purposes only, they are not exact laws.
They are true for most real systems, but counterexamples can
always be found \cite{schul1,schul2}. Also, they both may seem intuitively
evident, but to prove them rigorously is very difficult.
For ergodicity the relevant rigorous result
is the KAM (Kolmogorov–--Arnold--–Moser) theorem,
while for the destruction of the opposite time arrows
a rigorous theorem is still lacking.

Our results also resolve the ``contradiction" between the Prigogine's
``New Dynamics" \cite{prigogine1} (discussed in Sec.~\ref{SECcoars} of the present paper)
and Bricmont's comments \cite{bricmont}.
Dynamics of interacting systems can be divided into two types of
dynamics:
\begin{enumerate}
\item Reversible { \em ideal dynamics} is considered with respect
to the coordinate time, in which case entropy can either decrease
or increase. 
\item Irreversible { \em observable dynamics} is
considered with respect to the intrinsic time arrows of
interacting systems, in which case entropy increases as we can see
above. 
(In other words, the entropy increases rather than decreases with $t$ because
the orientation of $t$ is {\em defined} as a direction in which 
entropy increases.)
\end{enumerate}
In the framework of this terminology, the Prigogine's ``New Dynamics" \cite{prigogine1}
is one of the forms of the observable dynamics, while the Bricmont's paper \cite{bricmont} considers ideal dynamics. In particular, the observable
dynamics does not include Poincare's returns and reversibility, that are indeed unobservable by
a real observer, which makes it simpler than ideal dynamics. Yet, in principle, both
types of dynamics are correct.

This reasoning can also be applied to the interpretation of numerical results
in Sec.~\ref{SECnumsim}. Both Fig.~\ref{fignoint} and Fig.~\ref{figint} show 
that the arrow of time, defined by the total entropy, reverses at some point.
This reversal corresponds to the description by ideal dynamics.
But can such a reversal be observed? For the sake of conceptual clarity
we stress that the observer is nothing but one of the subsystems, and split the question into
three different ones. First, can the observer observe the reversal
of his {\em own} arrow of time? Second, can the observer observe the reversal of time arrow of 
his {\em total environment}? Third, can the observer observe the reversal of time arrow 
of a {\em small part} of his environment? 

The answer to the first question is {\em no}, because the observer naturally defines
the arrow of time as the direction in which his own entropy increases.
Namely, the observer perceives a subjective experience of the flow of time
because at each time he has access to his memory on some events which are not happening right now.
Such events are naturally interpreted as ``past'' by him. It can be argued that
memory can only work when entropy increases in the direction from the memorized event
to the time of recalling it (see, e.g., \cite{nik1}). Similarly, other processes in the brain
(or a computer) also seem to require an increase of entropy for their normal way of functioning 
(see also \cite{penrose}). In this way, one expects that the observer's subjective
experience of the flow of time always coincides with the direction in which
the observer's entropy increases. 

The answer to the second question is {\em almost certainly no}, because
if the total environment is observed then it interacts with the observer, 
and consequently their time arrows tend to be aligned, except, perhaps,
during a very short time needed for the process of aligning.

The answer to the third question is {\em sometimes yes, but usually no}.
Namely, some special systems (e.g., the spin-echo system discussed in Sec.~\ref{SEC2})
can weakly interact with its environment and still retain a time arrow opposite to that of its environment
for a relatively long time. Such special systems are relatively small parts of the total environment,
and the observer can observe that such a subsystem has a time arrow opposite 
to that of the observer. Indeed, as we have already explained in  Sec.~\ref{SEC2}, 
our results based on probabilistic reasoning on typical systems do not imply that
it is absolutely impossible to observe a subsystem in which entropy decreases.
They only explain why such systems are expected to be very rare, which agrees with everyday
experience.


A closely related observation is that
our results are not in contradiction
with the existence of dissipative systems \cite{prigogine2}
(such as certain self-organizing biological systems)
in which entropy of a subsystem can decrease with time, despite the fact
that entropy of the environment increases.
The full-system entropy (including the entropies of both the dissipative system and the environment)
increases, which is consistent with the entropy-increase law.
For such systems, it is typical that the interaction with the environment
is {\em strong}, while results of our paper refer to {\em weak}
interactions between the subsystems.
For example, for existence of living organisms, a strong energy
flow from the Sun is needed. The small flow from other stars is not sufficient for life, but is
sufficient for the decorrelation and for the alignment of the time arrows.
To quote from \cite{mac}: ``However, an observer is
macroscopic by definition, and all remotely interacting macroscopic systems become
correlated very rapidly (e.g. Borel famously calculated that moving a gram of material on the
star Sirius by 1 m can influence the trajectories of the particles in a gas on earth on a time
scale of $\mu$s \cite{borel})."

\section*{Acknowledgements}

The authors are grateful to the anonymous referees for 
various suggestions to improve the clarity of the paper.
The works of H.N. and V.Z. was supported by the Ministry of Science of the
Republic of Croatia under Contracts No.~098-0982930-2864 and 098-0352828-2863,
respectively.

\appendix

\section{Basic properties of the baker's map}
\label{app1}

In this appendix we present some basic properties of the
baker's map. More details can be found, e.g., in \cite{driebe}.

\subsection{Definition of the baker's map}

Consider a binary symbolic sequence
\begin{equation}
 \ldots S_{-2}, S_{-1}, S_0; S_1, S_2, S_3 \ldots
\end{equation}
infinite on both sides.
Such a sequence defines two real numbers
\begin{equation}
 x=0. S_1 S_2 S_3 \ldots , \;\;\;\;\; y=0. S_0 S_{-1} S_{-2} \ldots .
\end{equation}
The sequence can be moved reversibly with respect to the
semicolon in both directions. After the left shift we get new real numbers
\begin{equation}\label{eq27}
 x' = 2x - \lfloor 2x\rfloor , \;\;\;\;\; y'=\frac{1}{2} (y+\lfloor 2x\rfloor) ,
\end{equation}
where $\lfloor x\rfloor$ is the greatest integer less than or equal to
$x$. This map of unit square into itself is called the
{\em baker's map}.

\begin{figure*}[t]
\centerline{\includegraphics[width=15cm]{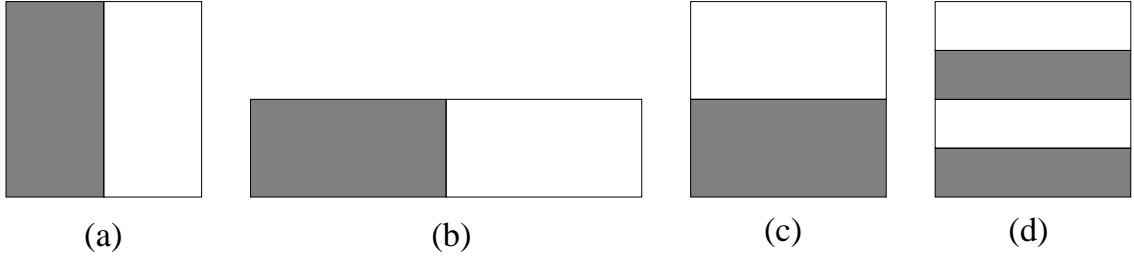}}
\caption{\label{fig1} Geometric interpretation of the baker's map. (a) Initial configuration.
(b) Uniform squeezing in vertical direction and stretching in horizontal direction
by a factor of 2. (c) The final configuration after cutting the right half and putting it over the left one. (d) The final configuration after two iterations.}
\end{figure*}

The baker's map has a simple geometrical interpretation presented in Fig.~\ref{fig1}.
There (a) is the initial configuration and (c) is the final configuration after one
baker's iteration, with an
intermediate step presented in (b). The (d) part represents the final configuration
after two iterations.

\subsection{Unstable periodic orbits}

The periodic symbolic sequences $(0)$ and $(1)$ correspond to fixed points $(x,y)=(0,0)$
and $(x,y)=(1,1)$, respectively. The periodic sequence $(10)$ corresponds to the period-2 orbit
$\{ (1/3,2/3), (2/3,1/3) \}$. From periodic sequence $\ldots 001;001 \ldots$
we get $\{(1/7,4/7), (2/7,2/7), (4/7,1/7) \}$. Similarly, from
$\ldots 011;011 \ldots$ we get $\{(3/7,6/7), (6/7,3/7), (5/7,5/7) \}$.

Any $x$ and $y$ can be approximated arbitrarily well by
$0.X_0\ldots X_n$ and $0.Y_0\ldots Y_m$, respectively, provided that
$n$ and $m$ are sufficiently large. Therefore the periodic sequence
$(Y_m\ldots Y_0 X_0\ldots X_n)$ can approach any point of the unit square
arbitrarily close. Thus, the set of all periodic orbits makes a dense set
on the unit square.

\subsection{Ergodicity, mixing, and area conservation}

Due to stretching in the horizontal direction, all close points diverge exponentially
under the baker's iterations. In these iterations, a random symbolic sequence
approaches any point of the square arbitrarily close.
In general, such an ergodic property can be used
to replace the ``time'' average $\langle A\rangle$ by the ``ensemble'' average
\begin{equation}
\langle A \rangle = \sum_n A(x_n,y_n) = \int A(x,y) \, d\mu(x,y) =
\int A(x,y) \rho(x,y) \, dx \, dy,
\end{equation}
where $d\mu(x,y)$ is the invariant measure and $\rho(x,y)$
is the invariant density for the map.
For the baker's map, $\rho(x,y)=1$.

Under the baker's iterations, any region maps into a set of narrow
horizontal strips. Eventually, it fills uniformly the whole unit square,
which corresponds to mixing. Similarly, reverse iterations
map the region into narrow vertical strips, which also corresponds to mixing.

During these iterations, the area of the region does not change.
This property is the area conservation law for the baker's map.

\subsection{Lyapunov exponent, shrinking and stretching directions}

If $x_0^{(1)}$ and $x_0^{(2)}$ have equal $k$ first binary digits, then,
for $n<k$,
\begin{equation}
 x_n^{(2)}-x_n^{(1)}=2^n (x_0^{(2)}-x_0^{(1)})
= (x_0^{(2)}-x_0^{(1)}) e^{n\log 2} ,
\end{equation}
where $\Lambda = \log 2$ is the first positive Lyapunov exponent for the baker's map.
Consequently, the distance between two close orbits increases exponentially
with increasing $n$, and after $k$ iterations becomes of the order of $1$.
This property is called sensitivity to initial conditions. Due to this property,
all periodic orbits are unstable.

Since the area is conserved, the stretching in the horizontal direction discussed above
implies that some shrinking direction must also exist. Indeed, the evolution
in the vertical $y$ direction is opposite to that of the horizontal
$x$ direction. If $(x_0^{(1)},y_0^{(1)})$ and $(x_0^{(2)},y_0^{(2)})$
are two points with $x_0^{(1)}=x_0^{(2)}$, then
\begin{equation}
 y_n^{(2)}-y_n^{(1)}=2^{-n} (y_0^{(2)}-y_0^{(1)})
= (y_0^{(2)}-y_0^{(1)}) e^{n(-\log 2)} .
\end{equation}
Hence $\Lambda = -\log 2$ is the second negative Lyapunov exponent for the baker's map.

\subsection{Decay of correlations}
\label{SECdecor}

Since $x$-direction is the unstable direction, the evolution in that direction
exhibits a decay of correlations.
The average correlation function $C(m)$ for a sequence $x_k$ is usually defined as
\begin{equation}
 C(m)=\lim \limits_{n \to
\infty}\frac{1}{n}\sum_{k=1}^n\left(x_k-\langle x\rangle\right) \left(x_{k+m}-\langle
x\rangle\right),
\end{equation}
where $\langle x\rangle=\lim \limits_{n \to\infty}\sum_{k=1}^n x_k/n$. Correlations can be more easily calculated if one knows the invariant measure $\mu(x)$, in which case
\begin{equation}
 C(m)=\int\left(x-\langle x\rangle\right)\left(f^m(x)-\langle x\rangle\right)d\mu(x),
\end{equation}
where $f^m(x)=x_m$ is the function that maps the variable $x$ to its image after $m$ iterations
of the map.
For the baker's map $d\mu(x)=dx$, so we can write
\begin{equation}
 C(m)=\sum_{j=0}^{2^m-1}\int_{j2^{-m}}^{(j+1)2^{-m}}\left(x-\langle
x\rangle\right)\left(2^mx-j-\langle
x\rangle\right)dx,
\end{equation}
which yields
\begin{equation}
C(m)=\sum_{j=0}^{2^m-1}\left[2^m\frac{x^3}{3}-\left(2^m\langle
x\rangle+\langle
x\rangle\right)\frac{x^2}{2}+\langle
x\rangle^2x-j\left(\frac{x^2}{2}-\langle
x\rangle x\right)\right]_{j2^{-m}}^{(j+1)2^{-m}}.
\end{equation}
For the baker's map $\langle x\rangle=1/2$, so
the sum above can be calculated explicitly
\begin{equation}
 C(m)=\frac{2^{-m}}{12}.
\end{equation}
This shows that the correlations decay exponentially with $m$.
The Pearson correlation for the system is given by
\begin{equation}
 r(m)=C(m)/C(0)=2^{-m}.
\end{equation}

\end{document}